\documentclass[aps,prb,twocolumn,showpacs,floatfix]{revtex4-1}

\usepackage{graphicx}
\usepackage{natbib}
\usepackage{amsmath}
\usepackage{color}

\newcommand{\uta}{Universit\'e de Toulouse; INSA-CNRS-UPS, LPCNO,
 135, Av. de Rangueil, 31077 Toulouse, France}
\newcommand{\uam}{Departamento de Ciencias B\'asicas,
 Universidad Aut\'onoma  Metropolitana-Azcapotzalco,
 Av. San Pablo 180, Col. Reynosa Tamaulipas, M\'exico D.F., Mexico}
\newcommand{\cnrs}{CNRS-LPN, Route de Nozay, 91460 Marcoussis, France}
\newcommand{\unam}{Instituto de F\'isica, Universidad Nacional Aut\'onoma
 de M\'exico, Apartado Postal 20-364,
 M\'exico Distrito Federal 01000, Mexico}

\newcommand{\spro}{\boldsymbol{\sigma}}
\newcommand{\ipro}{\iota}
\newcommand{\sop}{\hat{\boldsymbol{\sigma}}}
\newcommand{\sopel}{\hat{\sigma}}
\newcommand{\iop}{\hat{\iota}}

\begin{document}

\title{Room temperature optical manipulation of nuclear spin
       polarization in GaAsN
}

\author{C. Sandoval-Santana$^{1}$
    A. Balocchi$^{2}$,
           T. Amand$^{2}$,
        J. C. Harmand$^{3}$,
        A. Kunold$^{4}$,
        X. Marie$^{2}$}
\affiliation{$^1$\unam\\ $^2$\uta\\ $^3$\cnrs \\$^4$\uam}

\begin{abstract}
The effect of hyperfine interaction on the room-temperature defect-enabled spin
filtering effect in GaNAs alloys is investigated both experimentally and theoretically through a
master equation approach based on the hyperfine and Zeeman interaction between
electron and nuclear spin of the Ga$^{2+}_i$ interstitial spin filtering defect . We show that the nuclear
spin polarization of the gallium defect can be tuned through the optically induced spin polarization of conduction band electrons. 
\end{abstract}

\maketitle

\section{introduction}
Many applications such as quantum registers~\cite{hanson:161203,robledo:574}, quantum memories and nanoscale magnetic imaging setups~\cite{grinolds:1745}, rely on individually addressable spin systems that can be initialized and read out. 
Diamond NV paramagnetic centers~\cite{davies:1653} have provided with an ideal system to build such
applications given that its spin has a long coherence time that persists even at room temperature\cite{kennedy:4190}.
It has been shown that the spin states of NV centers can be coherently controlled by optical and radio-frequency means.
Similarly, interstitial Ga$^{2+}_i$ defects in dilute nitride GaAsN give rise to paramagnetic centers~\cite{wang:198} whose well isolated and stable spin
states can be addressed collectively and moreover detected both by optical and electrical means~\cite{zhao:241104,kunold:165202}. 
These defects, whose density can be controlled, are in particular responsible for the very high spin polarization of conduction band electrons in GaAsN compounds at room
temperature under circularly polarized light excitation~\cite{egorov:013539,kalevich:455,lombez:252115,kalevich:174} thanks to a very efficient spin filtering mechanism.  Recent measurements of an increased efficiency of the spin-filtering mechanism under a weak magnetic field in Faraday configuration
in GaAsN~\cite{kalevich:035205,puttisong:2013}  and in focus ion beam implanted InGaAs layers~\cite{nguyen:2013} have stimulated renewed interest in the subject after the early observations of D. Paget in GaAs~\cite{paget:931}.\\
Mainly three mechanisms have been proposed to understand the magnetic field induced amplification of the spin-filtering mechanism, all considering it as  a specific signature of the interplay between the Ga defect nucleus and its localized electron coupled through the hyperfine interaction
(HFI).\\
Paget\cite{paget:931} suggested that the electron spin polarization should increase in an external magnetic field as a result of the decoupling between the mixed electron and nuclear spins as calculated initially by Dyakonov and Perel~\cite{dyakonov:3059,lee:265}.
Kalevich \textit{et al.}~\cite{kalevich:035205} have interpreted the Faraday field enhancement of the 
spin filtering effect in GaAsN as due to a suppression of the chaotic magnetic field produced by the nuclei spin fluctuations surrounding each paramagnetic center, as typically observed in quantum dots~\cite{Braun_PRL,Petta_science}. These authors observed as well a shift of the photoluminescence (PL) polarization as a function of the applied magnetic field with respect to the zero field case, with opposite values for opposite excitation light helicity. This effect, explained in terms of the  dynamical polarization of the lattice nuclei (Overhauser effect) has been  phenomenologically introduced in the two-charge model~\cite{ivchenko:465804,weisbuch:141,kalevich:455,kalevich:208,kalevich:174} through a magnetic field dependence of the localized electron spin relaxation time. The model correctly reproduces the measured effects to what concerns the band-edge PL enhancements,  however it does not predict the observed shifts of the PL intensity versus magnetic field.\\
Puttisong \textit{et al.}~\cite{puttisong:2013} performed an analysis of the electron spin state mixing at the Ga$_i^{2+}$ as well considering the HFI and Zeeman contributions. They investigated the influence of HFI by low temperature (T=3 K) optically detected magnetic resonance experiments on the observed amplification of spin dependent recombination measured at room temperature. The model, centered on the HFI Hamiltonian,  focuses on the coupling between the localized electron and the Ga$^{2+}_i$ interstitial defect and compares the observed spin dependent recombination ratio (SDR$_r$) dependence on the magnetic field to the percentage of localized electron pure spin states in the paramagnetic centers. No zero magnetic field shift of the PL polarization curves is experimentally observed nor it can be 
predicted by the model.\\
Despite the success in describing the main features of the spin-filtering enhancement in a magnetic field, the different models proposed do not take into consideration 
the dynamics of the ensemble of the dilute nitride system composed of the conduction band electrons with its spin dependent recombination into the coupled nuclear-localized electron complex [inset of Fig. 1 (b)].\\
The aim of this paper is to present a comprehensive theoretical  work on the SDR related phenomena in GaAsN taking into consideration 
this whole dynamical system.
In order to gain insight into the interplay of the different mechanisms involved in the conduction band (CB) electron spin polarization in GaAsN alloys we develop a model based on the open quantum system approach that reduces to the well known two charge model\cite{ivchenko:465804,weisbuch:141,kalevich:455,kalevich:208,kalevich:174} in
the absence of HFI. This master equation model, taking into account the strong coupling of the center-localized electron and the gallium defect nuclei by
 hyperfine interaction~\cite{dyakonov:995},  is able to reproduce accurately all the observed features of the experimental results including the polarization dependent PL polarization shift in Faraday configuration. We show that on one hand, the HFI of the Ga$^{2+}_i$ centers causes strong mixing of the localized electron and nuclear spin states thus polarizing the nuclear spin and partially canceling the localized electron spin polarization. Under a magnetic field aligned parallel to the incident light, the coupling between electrons and nuclei is destroyed and electrons recover their spin polarization. On the other hand, the dynamical equilibrium of the HFI-coupled electron-nucleus eigenstates' populations under circularly polarized light leads to the appearance of an excitation power and polarization dependent shift of the electrons (conduction and localized) and paramagnetic center nuclei polarizations versus magnetic field in Faraday geometry with respect to the $B$=0 case.
From a macroscopic photoluminescence or photoconductivity polarization measurement it is then possible to deduce the average Ga$^{2+}_i$ interstitial electronic and nuclear spin polarizations and their evolution in a magnetic field. Though optical pumping of nuclear spins in semiconductors usually require cryogenic temperatures of the sample~\cite{meier:1984,dyakonov_spin_2008,urbaszek_nuclear_2013}, we show here that the nuclear spin states of an ensemble of Ga centers can be controlled and measured at room temperature through the spin polarization of conduction band electrons.\\
This paper is organized as follows. In Sec.~\ref{samples} we describe the
sample preparation, the experimental setup and present the experimental results.
The master equation model taking into account the hyperfine interaction
between the centers localized electrons and nuclei is described in
Sec. III.  The calculations and comparison with the experiment
are presented in Sec.~\ref{results}. Here, through the master equation approach, we demonstrate how
the hyperfine interaction significantly alters the spin polarization of conduction
band electrons, localized electrons and nuclei. We describe the mechanism behind the spin transfer from conduction band electrons to center's nuclei and the origin of the polarization shift under a Faraday magnetic field.
A summary of the results and the conclusions are drawn in Sec.~\ref{conclusions}.

\section{Samples and experimental results}\label{samples}
\begin{figure}
\includegraphics[width=0.5\textwidth]{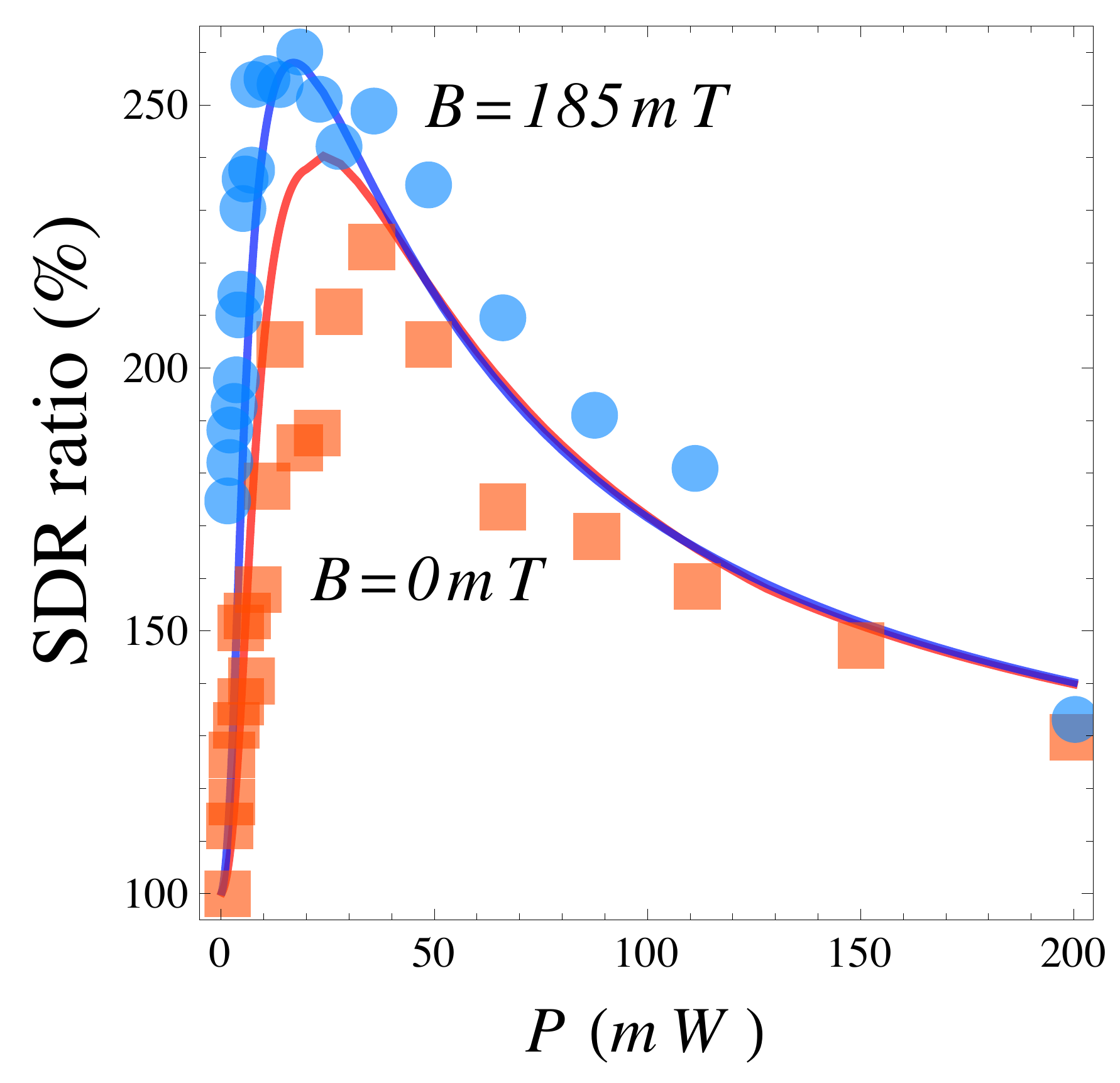}
\includegraphics[width=0.5\textwidth]{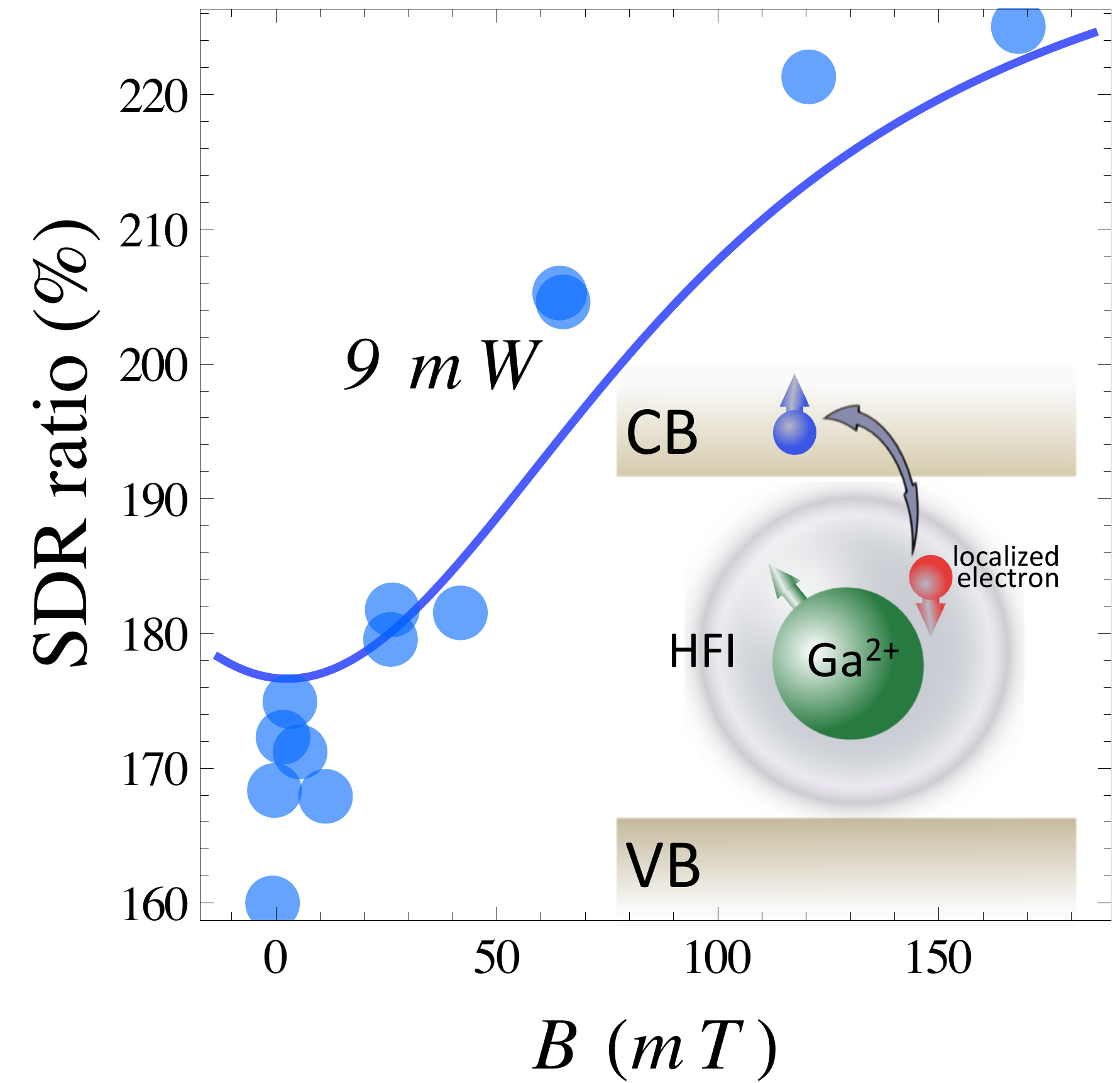}
\caption{(color on line) (a) Photoluminescence SDR$_r$ as a function of laser
excitation power $P$. The symbols show the experimental results; the solid lines indicate
the theoretical results for two different values of the magnetic field.
(b) Photoluminescence SDR$_r$ as a function of Faraday configuration
magnetic field for a fixed laser irradiance $P$=9 mW.
The circles indicate the experimental results while the solid line traces the
theoretical results under the same conditions. Inset: schematic representation of the SDR system showing the Ga$^{2+}_i$ atom with its localized electron coupled by the hyperfine interaction (HFI) and the photogenerated conduction band electron.}
\label{figure1}
\end{figure}
The sample under study consists  of a 100 nm thick GaAs$_{1-x}$N$_x$  layer (x=0.0079)  grown by molecular beam epitaxy on a (001)  
semi-insulating GaAs substrate  and capped with 10 nm GaAs.  The conduction band electron spin polarization properties in the structure have been investigated at room temperature by optical orientation experiments which rely on the transfer of the angular momentum of the exciting photons, using circularly polarized light, to the photogenerated electronic excitations. The excitation source is a continuous wave (CW) Ti:Sapphire laser emitting at 850 nm and focused onto the sample to a 50
$\mu$m diameter spot.  The excitation laser is either circularly ($\sigma^+$) or linearly ($\sigma^X$) polarized propagating along the $z$ growth axis
 and the resulting PL circular polarization ($P_c$) and SDR$_r$ are calculated respectively as $P_c = (I^{++} - I^{+-})/(I^{++} + I^{+-})$ and SDR$_r$=$I^+/I^X$. For calculating the PL circular polarization $P_c$, 
$I^{++}$ and $I^{+-}$ represent the PL intensity components co- and counter-polarized to the $\sigma^+$  excitation light. In the case of the SDR$_r$, $I^+$ and $I^X$
denote respectively the total PL intensities detected under a circular or linear excitation of same intensity.
The photoluminescence intensity is detected using a silicon photodiode coupled to a long-pass filter in order to suppress the contribution due to the laser scattered light and GaAs substrate/buffer layers luminescence. In order to improve the signal to noise ratio, the excitation light is mechanically chopped and the photodiode signal synchronously detected with a lock-in amplifier.\\
In Fig.~\ref{figure1} (a), squares, we present the photoluminescence SDR$_r$ as a function
of the laser excitation power $P$. We observe the main characteristic of the SDR effect, namely a marked excitation power dependence showing a peak value around $P$=25 mW~\cite{kunold:165202}. Figure~\ref{figure1} (a),  circles, reports the same experiment under a longitudinal magnetic field (Faraday configuration)  $B_z$=185 mT. As previously reported by Kalevich~\cite{kalevich:035205} we observed a sizable increase of the SDR ratio which is more substantial at low excitation power and gradually disappears at higher excitations. Figure~\ref{figure1} (b), reports the SDR ratio dependence on the Faraday magnetic field measured at $P$=9 mW. Altough a monotonous increase of the SDR$_r$ is observed~\cite{kalevich:035205}, the minimum SDR is observed for $B$=0 ; no shift is detected here in contrast to ref.[\!\!\citenum{kalevich:035205}] probably due to the low excitation power and the limited signal/noise ratio.\\
In order to account for our experimental observations and the additional evidences reported by Kalevich \textit{et al.}~\cite{kalevich:035205}, namely the shift from $B$=0 of the CB polarization dip under a magnetic field in Faraday geometry, we have developed in the next section a density matrix model comprising the
full dynamical system of CB electrons spin dependent recombination and  hyperfine interaction between localized electron and paramagnetic centers.
\begin{figure}
\includegraphics[width=8.00 cm]{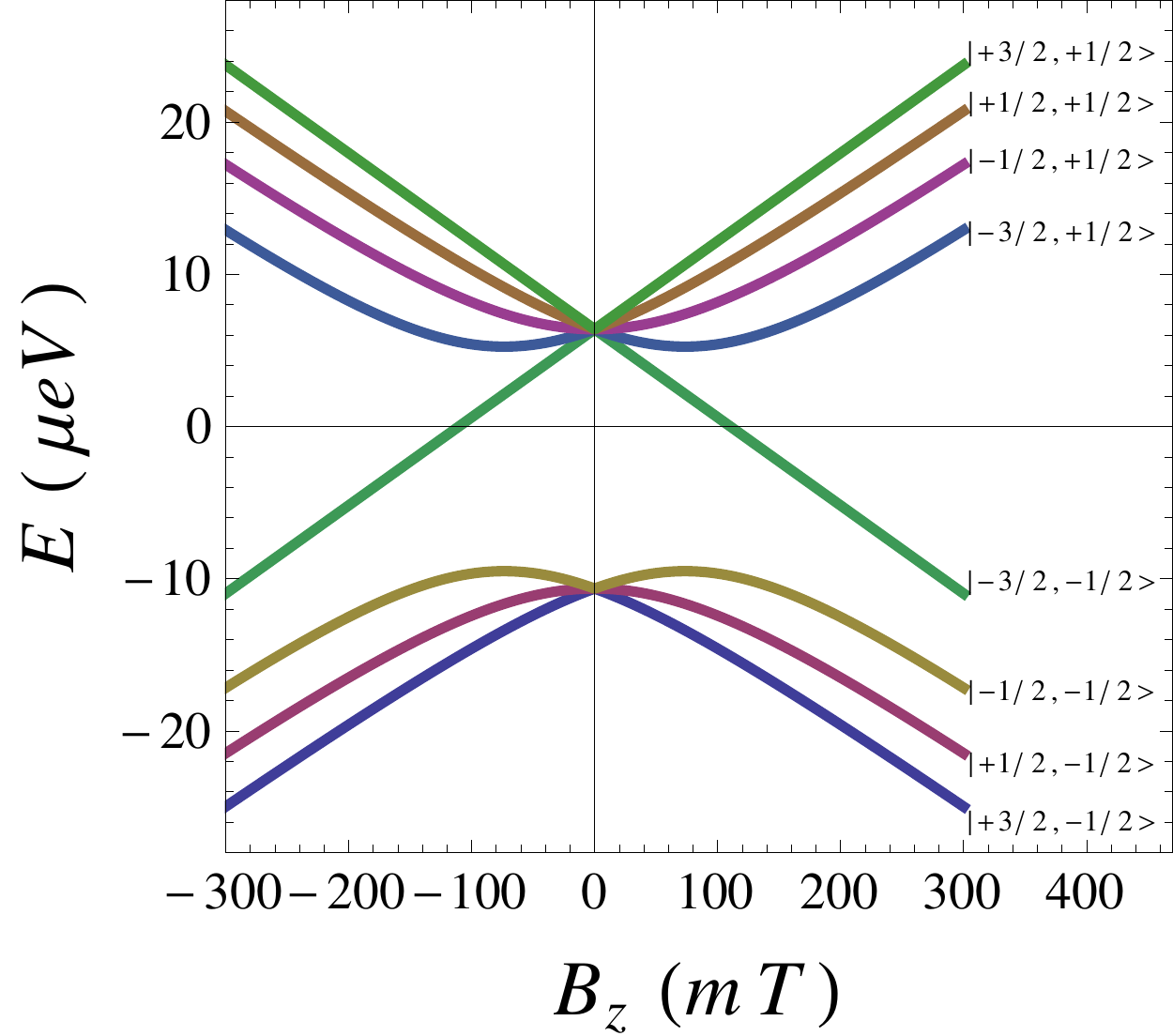}
\caption{(color on line) The calculated coupled Ga$^{2+}_i$ nuclei and localized electron spin state
structure as a result of hyperfine and Zeeman interactions. As the magnetic
field increases the nuclear and localized electron states decouple. The
decoupled spin states are presented at the right of the diagram. }
\label{figure2}
\end{figure}
\section{Model}\label{model}
\subsection{Two charge model}\label{model}
Our starting point is the two charge model based on the following set of rate equations:~\cite{ivchenko:465804,weisbuch:141,kalevich:455,kalevich:208,kalevich:174}
\begin{align}
\dot{n}
  &+\frac{\gamma_e}{2}\left(nN_1-4\boldsymbol{S}\cdot\boldsymbol{S}_c\right)=G,
  \label{mag:eq3}\\
\dot{p}&+\gamma_hN_2p=G,\\
\dot{N_1}
  &+\frac{\gamma_e}{2}\left(nN_1-4\boldsymbol{S}\cdot\boldsymbol{S}_c\right)
  -\gamma_hN_2p=0,\\
\dot{N_2}
  &-\frac{\gamma_e}{2}\left(nN_1-4\boldsymbol{S}\cdot\boldsymbol{S}_c\right)
  +\gamma_hN_2p=0.\label{mag:eq6}\\
\dot{\boldsymbol{S}}
  &+\frac{\gamma_e}{2}\left(\boldsymbol{S}N_1-\boldsymbol{S}_cn\right)
  +\frac{1}{\tau_s}\boldsymbol{S}+\boldsymbol{S}\times\boldsymbol{\omega}
  =\boldsymbol{\Delta G},\label{mag:eq1}\\
\dot{\boldsymbol{S}}_c
  &+\frac{\gamma_e}{2}\left(\boldsymbol{S_c}n-\boldsymbol{S}N_1\right)
  +\frac{1}{\tau_{sc}}\boldsymbol{S_c}
  +\boldsymbol{S_c}\times\boldsymbol{\Omega}
  =\boldsymbol{0}.\label{mag:eq2}
\end{align}
Here $n$ is the density of conduction band (CB) electrons,
the total number of unpaired paramagnetic traps is given by
$N_1$ and $N_2$ is the total number of electrons singlets
hosted by the paramagnetic traps. $\boldsymbol{S},\boldsymbol{S_c}$ represent the average free and localized unpaired total electron spin.
  Holes ($p$) are considered unpolarized\cite{hilton:146601}
as their spin relaxes with a characteristic time of the order of 1 ps~\cite{malinowski}.
Eqs. (\ref{mag:eq3})-(\ref{mag:eq2}) ensure conservation of charge
neutrality and number of centers:
\begin{eqnarray}
n-p+N_2=0\label{chgcons:eq1},\\
N_1+N_2=N\label{cencons:eq1}.
\end{eqnarray}
The terms of the form $-\gamma_e\left(nN_1-4\boldsymbol{S}\cdot\boldsymbol{S}_c\right)/2$
and $\gamma_e\left(\boldsymbol{S}_cn-\boldsymbol{S}N_1\right)/2$
are responsible for the
spin dependent free electron capture in paramagnetic centers with recombination
rate $\gamma_e$.
The recombination rate of conduction electrons to paramagnetic centers
is increased when the free and localized electrons total spins $\boldsymbol{S}$ and $\boldsymbol{S}_c$ are antiparallel whereas
it vanishes when they are parallel as expected from the Pauli exclusion
principle needed to form a singlet state [inset of Fig. 1(b)]. The terms $-\gamma_hpN_2$ model the
spin independent recombination of one electron of the paramagnetic center
singlet with a hole. The photo generation of spin-up and spin-down electrons is accounted for by
the terms $G_{+}$ and $G_{-}$ and of holes by $G=G_++G_-$ using the same method as in Ref.[\!\!~\citenum{kunold:165202}].
In CW conditions the total photoluminescence intensity under linear ($X$)
or circular ($+$) excitation is calculated as
$I^{X(+)}=\gamma_r n\left(t\right)p\left(t\right)$ where 
 $t$ is a sufficiently
long time to ensure steady state conditions.
In the absence of the SDR mechanism,
$\mathrm{SDR}_r=1$ whereas in its presence $\mathrm{SDR}_r>1$.
Magnetic field effects such as the Hanle effect, are included into the model
via the spin precession terms that arise from the Zeeman interaction
$\boldsymbol{\omega}\times\boldsymbol{S}$ for free electrons and
$\boldsymbol{\Omega}\times\boldsymbol{S}_c$ for localized electrons
where $\boldsymbol{\omega}= g \mu_B \hbar\boldsymbol{B}$,
$\boldsymbol{\Omega}=g_c\mu_B \hbar\boldsymbol{B}$ and
$\mu_B$ is the Bohr magneton.
The gyromagnetic factors for free electrons and localized electrons
were set to $g$=1 and $g_c$=2 respectively~\cite{kalevich:174,kalevich:208,pettinari:245202,zhao:041911}.
Nevertheless, the inclusion of the localized electron-nucleus hyperfine interaction
terms in Eqs. (1)-(6) giving rise to the amplification of the SDR in longitudinal magnetic field 
is not straightforward as we shall see in the next section.

\subsection{Master equation}
When an electron is bounded to a deep Ga$^{2+}_i$ defect, its wavefunction is strongly localized~\cite{lagarde:208,wang:198,puttisong:2013}  and one can consider that its spin interacts mainly with the corresponding unique Ga nucleus yielding coupled electron-nucleus quantum states. Note that this is a very different situation compared to the usual treatment of hyperfine interaction of weakly localized (for instance electrons bound to donor states) or confined electrons (in quantum dots) in which the electron wavefunction interacts with 10$^{5}$ - 10$^{6}$ nuclei allowing a mean field description~\cite{meier:1984,dyakonov_spin_2008,urbaszek_nuclear_2013}.
The hyperfine interaction $A \boldsymbol{\hat{I}}\cdot \boldsymbol{\hat{S}}_c$ (where $A$, $\boldsymbol{\hat{I}}$ and $\boldsymbol{\hat{S}}_c$ are respectively the hyperfine interaction constant, the nucleus and localized electron spin operators) between the localized electron and the interstitial Ga$^{2+}_i$  in the Hamiltonian
gives rise to the following eigenstates in zero magnetic field
\begin{eqnarray}
\left\vert 1,1\right\rangle &=& -\frac{\sqrt{3}}{2}\left\vert \frac{3}{2},-\frac{1}{2}\right\rangle
                                       +\frac{1}{2}\left\vert +\frac{1}{2},+\frac{1}{2}\right\rangle,\\
\left\vert 1,0\right\rangle &=& -\frac{1}{\sqrt{2}}\left\vert +\frac{1}{2},-\frac{1}{2}\right\rangle
                                       +\frac{1}{\sqrt{2}}\left\vert -\frac{1}{2},+\frac{1}{2}\right\rangle, \\
\left\vert 1,-1 \right\rangle &=& -\frac{1}{2}\left\vert -\frac{1}{2},-\frac{1}{2}\right\rangle
                                       +\frac{\sqrt{3}}{2}\left\vert -\frac{3}{2},+\frac{1}{2}\right\rangle,\\      
\left\vert 2,2 \right\rangle &=& \left\vert + \frac{3}{2},+\frac{1}{2}\right\rangle,\\ 
\left\vert 2,1 \right\rangle &=&  \frac{1}{2}\left\vert +\frac{3}{2},-\frac{1}{2}\right\rangle 
                                          +   \frac{\sqrt{3}}{2}\left\vert +\frac{1}{2},+\frac{1}{2}\right\rangle ,\\
\left\vert 2,0 \right\rangle &=&  \frac{1}{\sqrt{2}}\left\vert +\frac{1}{2},-\frac{1}{2}\right\rangle 
                                          +   \frac{1}{\sqrt{2}}\left\vert -\frac{1}{2},+\frac{1}{2}\right\rangle,\\
\left\vert 2,-1 \right\rangle &=&  \frac{\sqrt{3}}{2}\left\vert -\frac{1}{2},-\frac{1}{2}\right\rangle 
                                          +   \frac{1}{2}\left\vert -\frac{3}{2},+\frac{1}{2}\right\rangle,\\ 
\left\vert 2,-2 \right\rangle &=&  \left\vert -\frac{3}{2},-\frac{1}{2}\right\rangle .
\end{eqnarray}
where $\left\vert j,j_z\right\rangle$ on the left hand side represent the eigenstate of total angular momentum $j$ and component $j_z$, whereas on the right hand side $\left\vert m,s\right\rangle$ is a state of the uncoupled nuclear spin $m$ and localized electron
spin $s$  projections along the chosen quantization axis.
The total Hamiltonian (taking into account the conduction electron and the coupled localized electron-Ga nucleus system) takes the following form: 
\begin{equation}\label{ham:eq1}
\hat{H}=\boldsymbol{\omega}\cdot\hat{\boldsymbol{S}}
 +\boldsymbol{\Omega}\cdot\hat{\boldsymbol{S}}_c
 +\boldsymbol{\Theta}\cdot\hat{\boldsymbol{I}}
 +A\hat{\boldsymbol{I}}\cdot\hat{\boldsymbol{S}}_c,
\end{equation}
where the first three terms correspond to the Zeeman interaction
between an external magnetic field $\boldsymbol{B}$ and the
magnetic moments of CB electrons, localized electrons and nuclei.
The spin precession terms in the rate equations arise from these
contributions.
In the last term, accounting for the hyperfine interaction
between the center's nuclei and the localized electrons, 
$A=687\times 10^{-4}$cm$^{-1}$ was set to the average hyperfine parameter\cite{wang:198}
of the two gallium isotopes: $^{69}$Ga$^{2+}_i$ (60\%) 
with $A=620\times 10^{-4}$cm$^{-1}$ and
$^{71}$Ga$^{2+}_i$ (40\%) with $A=788\times 10^{-4}$cm$^{-1}$.
Fig. 2 reports the calculated energies of the coupled localized electron- Ga$^{2+}_i$ nucleus states as a function of a Faraday magnetic field. For zero magnetic field the 
hyperfine interaction mixes the nucleus and electron spin states giving rise to two degenerate states corresponding to the two possible values of the total angular momentum $J=I+S$=2,1.  As an external magnetic field is applied in Faraday geometry, the mixing due to the hyperfine interaction progressively decreases as the 
electron Zeeman term becomes predominant. For sufficiently high magnetic field values the electron and nucleus
are effectively decoupled and pure electron and nuclear spin states are now the eigenstates of the system.\\
The form of the hyperfine interaction terms reveals the difficulties
of introducing its effects
directly into the rate Eqs. (1)-(6).
The Zeeman interactions are comprised only of CB and localized electrons angular momentum
linear terms.
Their corresponding angular momentum operators form a closed algebra, characterized by
$\left[S_i,S_j\right]=i\hbar \sum_{k=x,y,z}\epsilon_{ijk}S_k$ ($\epsilon_{ijk}$ is the Levi-Civita symbol), thus yielding
one time dependent differential equation for each component of the angular momentum
arising from the commutator in the von Neumann equation. Therefore
Eqs. (\ref{mag:eq1}) and (\ref{mag:eq2}) contain only linear terms in the angular momentum.
Unlike the Zeeman terms,  the hyperfine interaction in the
Hamiltonian (\ref{ham:eq1})  is the product
$\hat{\boldsymbol{I}}\cdot\hat{\boldsymbol{S}}_c$ that does not give rise to
a closed algebra. An attempt to workout the rate equations starting from the
Hamiltonian (\ref{ham:eq1}) would yield an increasingly
large number of differential equations.
Therefore, in this case the master equation formulation seems to be a better candidate to
model the SDR  than
the rate equation approach.\\
The master equation for the density matrix $\hat\rho$ for the given
system is thereby expressed as
\begin{equation}\label{mastereq}
\dot{\hat{\rho}}
=\frac{i}{\hbar}\left[\hat{\rho},\hat{H}\right]
+\mathcal{D}\left[\hat{\rho}\right]+\hat G,
\end{equation}
where the Hamiltonian $\hat{H}$ is given by (\ref{ham:eq1}) and
$\mathcal{D}\left[\hat{\rho}\right]$ is the dissipator.
Accordingly, the chosen basis is comprised of a non interacting  ensemble of $1/2$ spin
CB electrons, spin unpolarized valence band (VB) holes, spin polarized localized electrons, $3/2$ spin nuclei and paired
(singlet) localized electrons. The dissipator $\mathcal{D}\left[\hat{\rho}\right]$ describes the
coupling or decay channels resulting from an interaction with the
photon environment and spin relaxation
as  explained later on.
Similarly $\hat{G}$ corresponds to
the laser generating term. 
We can identify a total of 12 states:
i) one hole state,
ii) one paired localized state, 
iii) one spin down  and 
iv) one spin up CB electron state,
v) a total of eight states, Eqs. (9)-(16), corresponding to the localized electron-nucleus states, \textit{i.e.} two states for the spin up and
spin down localized electron times four states for the $3/2$ Ga$^{2+}_i$ nucleus spin. This basis is displayed explicitly in the Appendix.

In order to connect the master equation and the rate equation formulations
we must build the operators
corresponding to the ensemble averages $n$, $p$, $N_1$, $N_2$, $\boldsymbol{S}$,
$\boldsymbol{S}_c$ in the rate Eqs. (\ref{mag:eq3})-(\ref{mag:eq2}) and the nuclei
angular momentum operator $\hat{\boldsymbol{I}}$ as detailed in the Appendix.

Now we turn our attention to the explicit form of
the dissipator
$\mathcal{D}\left[\hat\rho\right]$.
It contains the interactions that lead to spin dependent
recombination between CB electrons and localized electrons;
the successive spin-independent recombination of localized electrons to the VB;
bimolecular recombination between CB and VB electrons and
finally CB, localized electrons and nuclei
spin decoherence and relaxation. In the absence of
hyperfine interaction its structure should permit to retrieve
the two charge model rate Eqs. (\ref{mag:eq3})-(\ref{mag:eq2}).
It is given by
\begin{multline}
\mathcal{D}\left[\hat\rho\right]=
-\left(\gamma_r p n+\gamma_h p N_2\right)\hat p
-\frac{1}{2}\left(\frac{\gamma_e}{2} Q+\gamma_r p n\right)\hat n\\
+\left(\frac{\gamma_e}{2} Q-\gamma_h pN_2\right)\hat N_2
-\frac{1}{8}\left(\frac{\gamma_e}{2}Q-\gamma_h p N_2\right)\hat N_1\\
-2\left[\frac{\gamma_e}{2}\left(N_1\boldsymbol{S}-n\boldsymbol{S}_c\right)
+\frac{1}{\tau_s}\boldsymbol{S}+\gamma_r p\boldsymbol{S}\right]
\cdot \hat{\boldsymbol{S}}-
\frac{\boldsymbol{I}\cdot\hat{\boldsymbol{I}}}{10\tau_n}\\
-2\sum_{i=-3/2}^{3/2}
\left[\frac{\gamma_e}{2}
\left(n \spro_i- \boldsymbol{S}\ipro_i\right)
+\frac{\spro_i}{\tau_{sc}}
\right]\cdot\sop_i .\label{dissipator}
\end{multline}
Here $n$, $p$, $N_1$, $N_2$, $\boldsymbol{S}$ and
$\boldsymbol{S}_c$ are the
variables introduced in Sec. \ref{model} and
$\hat n$, $\hat p$, $\hat{N}_1$, $\hat{N}_2$,
$\hat{\boldsymbol{S}}$ are the corresponding operators
whose explicit form is given in App. \ref{operators}.
We consider them to be connected through the
ensemble averages
$n={\rm Tr}\left[\hat n\hat \rho\right]$,
$p={\rm Tr}\left[\hat p\hat \rho\right]$,
$N_1={\rm Tr}\left[\hat{N}_1\hat \rho\right]$,
$N_2={\rm Tr}\left[\hat{N}_2\hat \rho\right]$,
$\boldsymbol{S}={\rm Tr}\left[\hat{\boldsymbol{S}}\hat \rho\right]$ and
$\boldsymbol{S}_c={\rm Tr}\left[\hat{\boldsymbol{S}}_c\hat \rho\right]$.
For the sake of brevity we have defined
\begin{equation}\label{qdef:eq1}
Q=n N_1-4\boldsymbol{S}\cdot \boldsymbol{S}_c.
\end{equation}
The terms proportional to $\gamma_e$ are the spin dependent recombination
rates of CB electrons recombining to the
paramagnetic traps. Localized electrons recombine to
the VB at a rate given by $\gamma_h$.
Those terms proportional to $\gamma_r$ are related to
bimolecular recombination.
Spin relaxation for CB and localized electrons is modeled by the terms
proportional to $1/\tau_s$  and $1/\tau_{sc}$ respectively.
We introduce a phenomenological nuclear spin decay term $\frac{\mathbf{I}\cdot \hat{\mathbf{I}}}{10\tau_n}$ to take into account 
possible mechanisms such as the fluctuating dipole-dipole interaction between the Ga interstitial with its neighbors, the fluctuating hyperfine interaction
with conduction electrons and also electron exchange between the center and the free conduction electrons\cite{} , these mechanisms arising when the center is occupied by an electron singlet.\\
We introduce the nuclear angular momentum operator
$\hat{\boldsymbol{I}}$ and its corresponding
ensemble average
$\boldsymbol{I}={\rm Tr }\left[\hat{\boldsymbol{I}}\rho\right]$.
The auxiliary operators $\sop_i$ and $\iop_i$ with
$i=-3/2$, $-1/2$, $1/2$, $3/2$ are also presented
in the Appendix. Their ensemble averages
are given by $\spro_i={\rm Tr}\left[\sop_i\rho\right]$ and
$\ipro_i={\rm Tr}\left[\iop_i\rho\right]$ respectively.
These operators are related to the localized electron number and
their angular momentum and therefore have the following properties
$\hat{\boldsymbol{S}}_c=\sum_{m=-3/2}^{3/2}\sop_m$
and $\hat N_1=\sum_{m=-3/2}^{3/2}\iop_m$.

The dissipator $\mathcal{D}\left[\hat\rho\right]$ in
Eq. (\ref{dissipator}) is constructed
as a linear combination of the elements of
the orthogonal inner product space
spanned by the set of operators $\mathcal{V}=
\{\hat n$, $\hat p$, $ \hat{N}_1$, $ \hat{N}_2$, $
\hat{S}_x$, $\hat{S}_y$, $\hat{S}_z$, $
\hat{S}_{cx}$, $ \hat{S}_{cy}$, $\hat{S}_{cz}$, $
\hat{I}_x$, $\hat{I}_y$, $\hat{I}_z$, $
\sopel_{xm}$, $\sopel_{ym}$, $\sopel_{zm}\}$.
The vector space $\mathcal{V}$ inner product is
conveniently set to be the trace of the product
of any two matrices belonging to $\mathcal{V}$.
Therefore if $\hat V_i$ and $\hat V_j$
are elements of $\mathcal{V}$ then
${\rm Tr}\left[\hat V_i\hat V_j\right]=
{\rm Tr}\left[\hat V_i^2\right]\delta_{ij}$.
Thus, for example, to obtain the dynamical equation
for CB electrons we first calculate $\dot n$ using
the master equation
\begin{multline}
\dot n={\rm Tr}\left[ \hat n\dot {\hat \rho}\right]=
{\rm Tr}\left\{
\frac{i}{\hbar}\hat n\left[\hat\rho,\hat H\right]+
\hat n \mathcal{D}\left[\hat\rho\right]+
\hat n\hat G
\right\}\\
={\rm Tr}\left\{
\frac{i}{\hbar}\left[\hat n,\hat H\right]\hat\rho+
\hat n \mathcal{D}\left[\hat\rho\right]+
\hat n\hat G
\right\},
\end{multline}
where the commutor in the last line of the previous
equation vanishes. Second,
we calculate the dissipator term by using the orthogonality
of the matrix vector inner space and the explicit form
of the dissipator (\ref{dissipator})
\begin{multline}
{\rm Tr}\left\{
\hat n \mathcal{D}\left[\hat\rho\right]
\right\}=
-\frac{1}{2}\left(\frac{\gamma_e}{2} Q+\gamma_r p n\right)
{\rm Tr}\left[\hat n^2\right]\\
=-\left(\frac{\gamma_e}{2} Q+\gamma_r p n\right).
\end{multline}
Finally we calculate the generating term part
${\rm Tr}\left[\hat n\hat G\right]=G$. Collecting these results
we retrieve the CB electron density Eq. (\ref{mag:eq3}). This procedure can be repeated
for Eqs. (\ref{mag:eq3})-(\ref{mag:eq2}). It is
important to stress that the obtained $\boldsymbol{S}_c$ rate
equations contain additional terms compared to the two charge model ones
arising from the hyperfine interaction.
Even though $\left(1/2\right)\left[-\left(\gamma_e/2\right)\left(\boldsymbol{S}_c n
-\boldsymbol{S}N_1\right)-\boldsymbol{S}_c/\tau_{sc}\right]
\cdot\hat{\boldsymbol{S}}_c$
would at first glance seem a simpler option for the
last term in Eq. (\ref{dissipator}), it does not
guarantee equal recombination rates from the CB electron
spin states to the four nuclear spin states that might lead
to negative density matrix probabilities in the
high power regime.
Instead, $-2\sum_{m=-3/2}^{3/2}
\left[(\gamma_e/2)
\left(n \spro_m- \boldsymbol{S}\ipro_m\right)
+\spro_m/\tau_{sc}
\right]\cdot\sop_m$ not only
yields uniform recombination rates for all nuclear
spin states but also reproduces the localized electrons
polarization rate equations as can be readily verified
by applying the orthogonality properties of the auxiliary operators.
To understand the amplification of the spin filtering effect observed in Fig. 1 under longitudinal magnetic field, it is
important to take into account the transfer of angular momentum between
CB electrons, localized electrons and traps. Using the master equation
(\ref{mastereq}) and the explicit form of the dissipator (\ref{dissipator})
we work out the total change in angular momentum as
\begin{multline}
\frac{d}{dt}\left(\boldsymbol{S}+\boldsymbol{S}_c+\boldsymbol{I}\right)
={\rm Tr}\left[\left(\hat{\boldsymbol{S}}+\hat{\boldsymbol{S}}_c+\hat{\boldsymbol{I}}\right)
\dot{\hat{\rho}}\right]\\
=-\frac{1}{\tau_s}\boldsymbol{S}-\frac{1}{\tau_{sc}}\boldsymbol{S}_c
-\frac{1}{\tau_n}\boldsymbol{I}+\boldsymbol{\Delta}\boldsymbol{G}\\
+\boldsymbol{\omega}\times\boldsymbol{S}+\boldsymbol{\Omega}\times\boldsymbol{S}_{c}
+\boldsymbol{\Theta}\times\boldsymbol{I}.
\end{multline}
Here it should be noted that no terms arising from the hyperfine coupling
contribute to the total angular momentum losses.
Under a magnetic field in Faraday configuration the last three terms in the
previous equations vanish, and under steady state conditions
\begin{equation}
\frac{1}{\tau_s}\boldsymbol{S}+\frac{1}{\tau_{sc}}\boldsymbol{S}_c
+\frac{1}{\tau_n}\boldsymbol{I}=\boldsymbol{\Delta}\boldsymbol{G}.\\
\end{equation}
Moreover, if we separate the angular momentum change in
the CB and localized electron part and the nuclear part we
obtain
\begin{eqnarray}
\frac{d}{dt}\left(\boldsymbol{S}+\boldsymbol{S}_c\right)
&=&-\frac{1}{\tau_s}\boldsymbol{S}-\frac{1}{\tau_{sc}}\boldsymbol{S}_c
+\boldsymbol{\omega}\times\boldsymbol{S}
+\boldsymbol{\Omega}\times\boldsymbol{S}_{c}\nonumber\\
&&+A{\rm Tr}\left[\hat{\boldsymbol{S}}_c\times\hat{\boldsymbol{I}}\hat\rho\right]
+\boldsymbol{\Delta}\boldsymbol{G},\label{momcons:eq1}\\
\frac{d}{dt}\boldsymbol{I}&=&-\frac{1}{\tau_n}\boldsymbol{I}
+\boldsymbol{\Theta}\times\boldsymbol{I}\nonumber\\
&&-A{\rm Tr}\left[\hat{\boldsymbol{S}}_c\times\hat{\boldsymbol{I}}\hat\rho\right]
\label{momcons:eq2},
\end{eqnarray}
where we observe that the hyperfine coupling term
$A{\rm Tr}\left[\hat{\boldsymbol{S}}_c\times\hat{\boldsymbol{I}}\hat\rho\right]$
 transfers angular momentum
from the CB-center system to the nucleus until steady state conditions are reached.
This effect can be in principle integrated together with the localized electron spin losses
by replacing it by a time dependent relaxation time $\tau_{sc}\left(B\right)$ as phenomenologically introduced in Ref. [\!\!~\citenum{kalevich:035205}].

\section{Results}\label{results}
\begin{figure}
\includegraphics[width=8.00 cm]{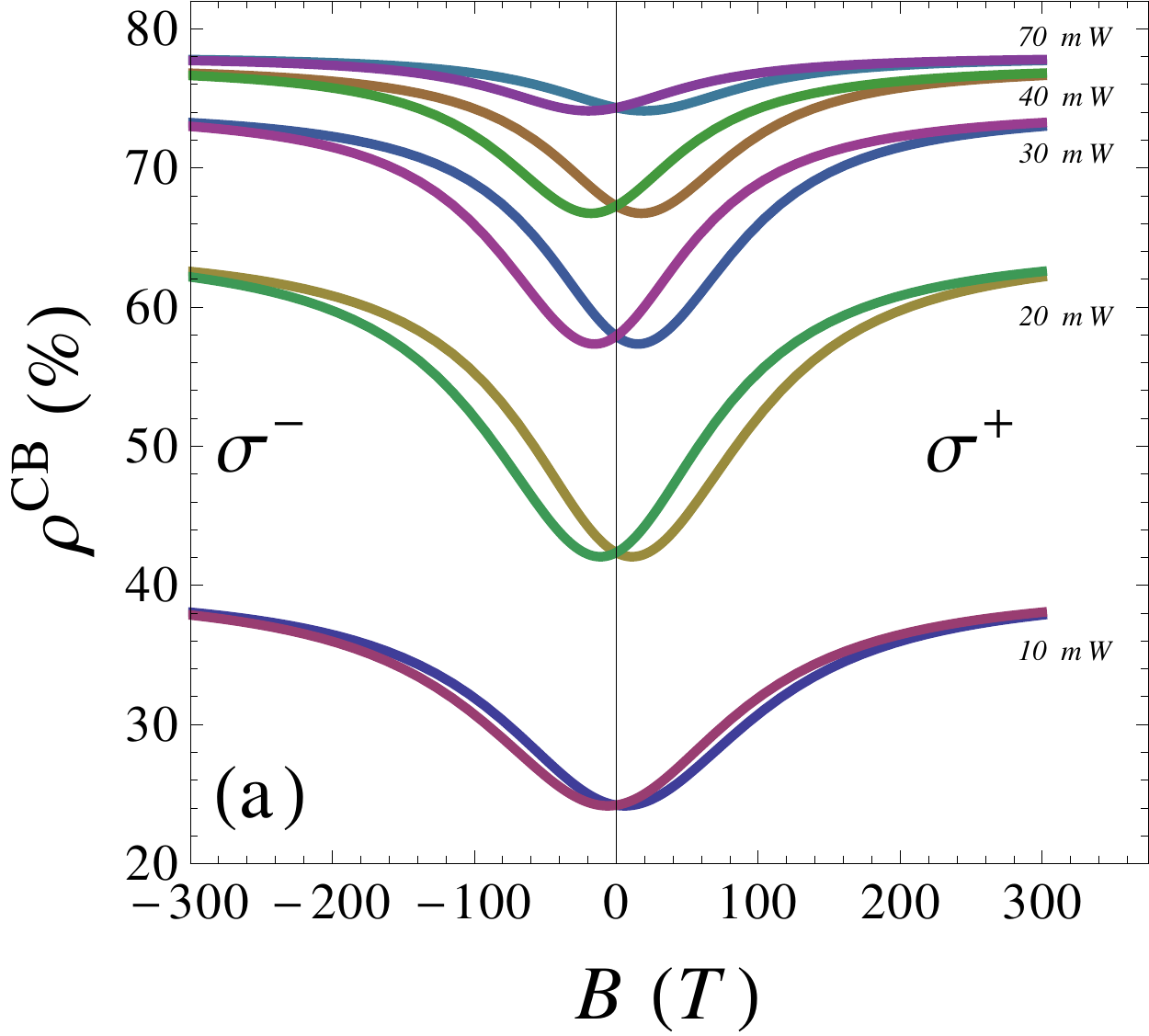}
\includegraphics[width=8.00 cm]{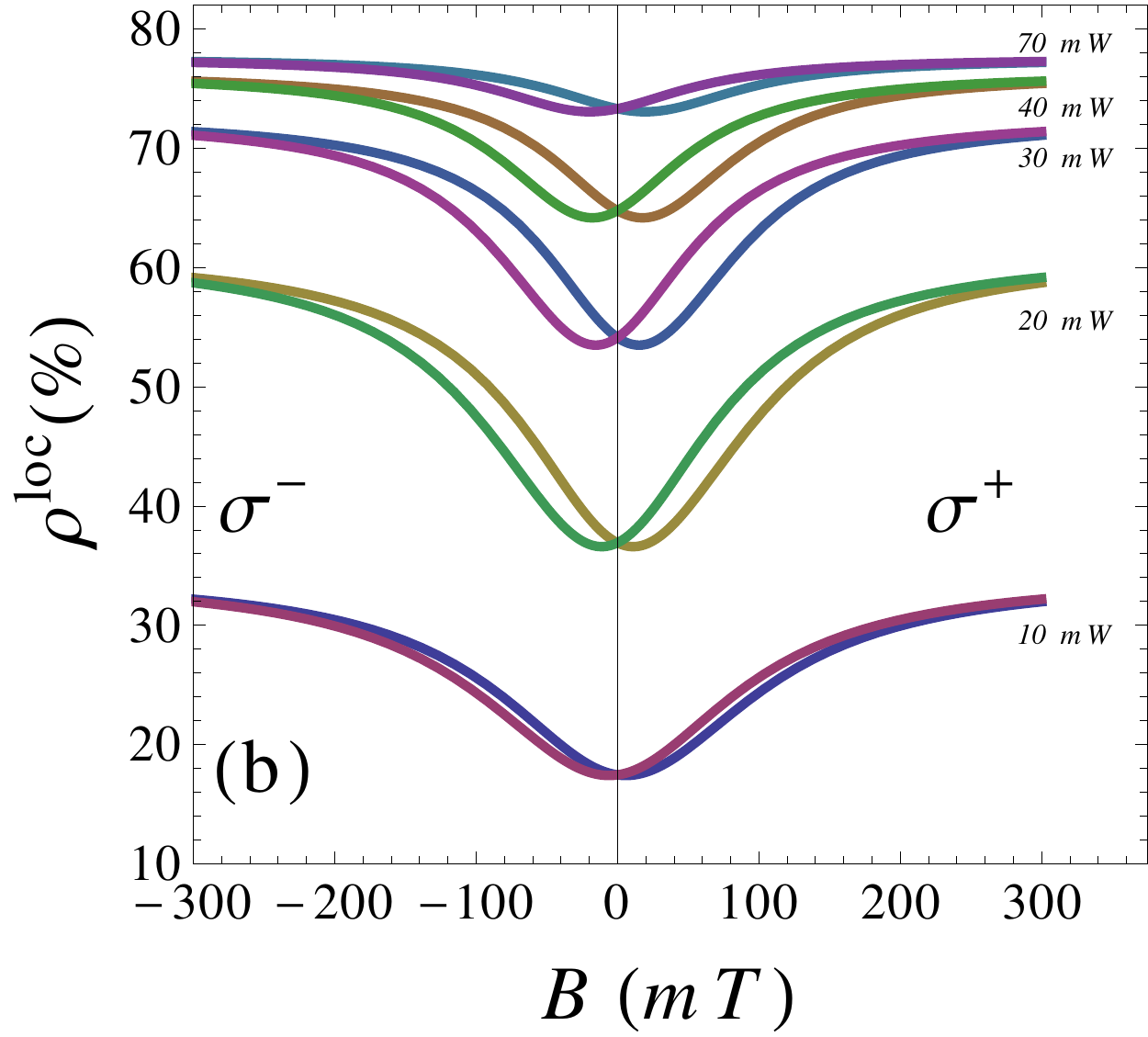}
\includegraphics[width=8.00 cm]{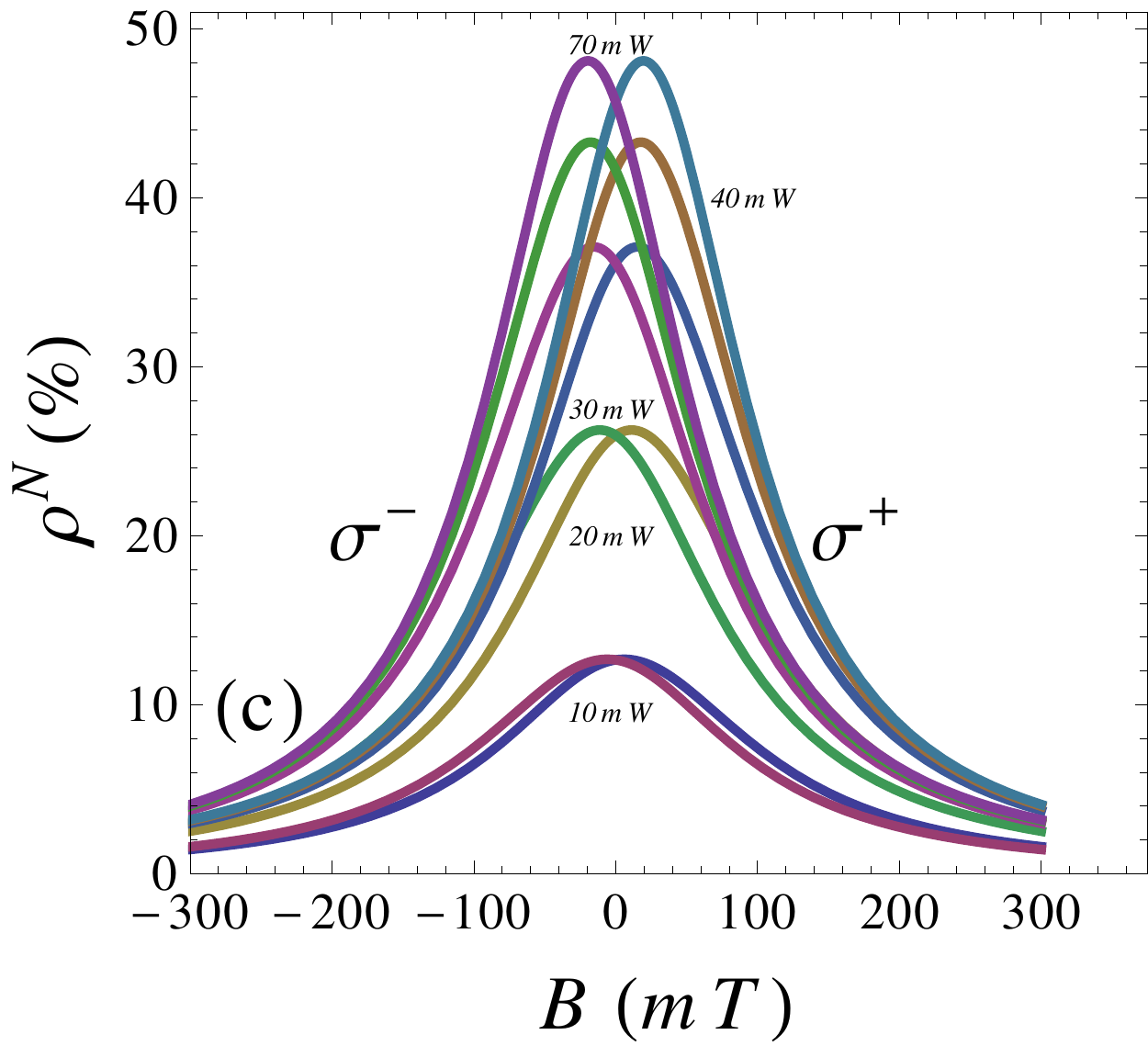}
\caption{(color on line) The calculated spin polarization degree of (a) CB electrons,
(b) localized electron and (c) Ga$^{2+}_i$ nuclei
as a function of the Faraday configuration magnetic field
for excitation powers from 10 to 70 mW and for right ($\sigma^+$) and left ($\sigma^-$) circularly
polarized light. The positive (negative) field extrema correspond to a $\sigma^+$($\sigma^-$) excitation.
}
\label{figure4}
\end{figure}
\begin{figure}
\includegraphics[width=8.00 cm]{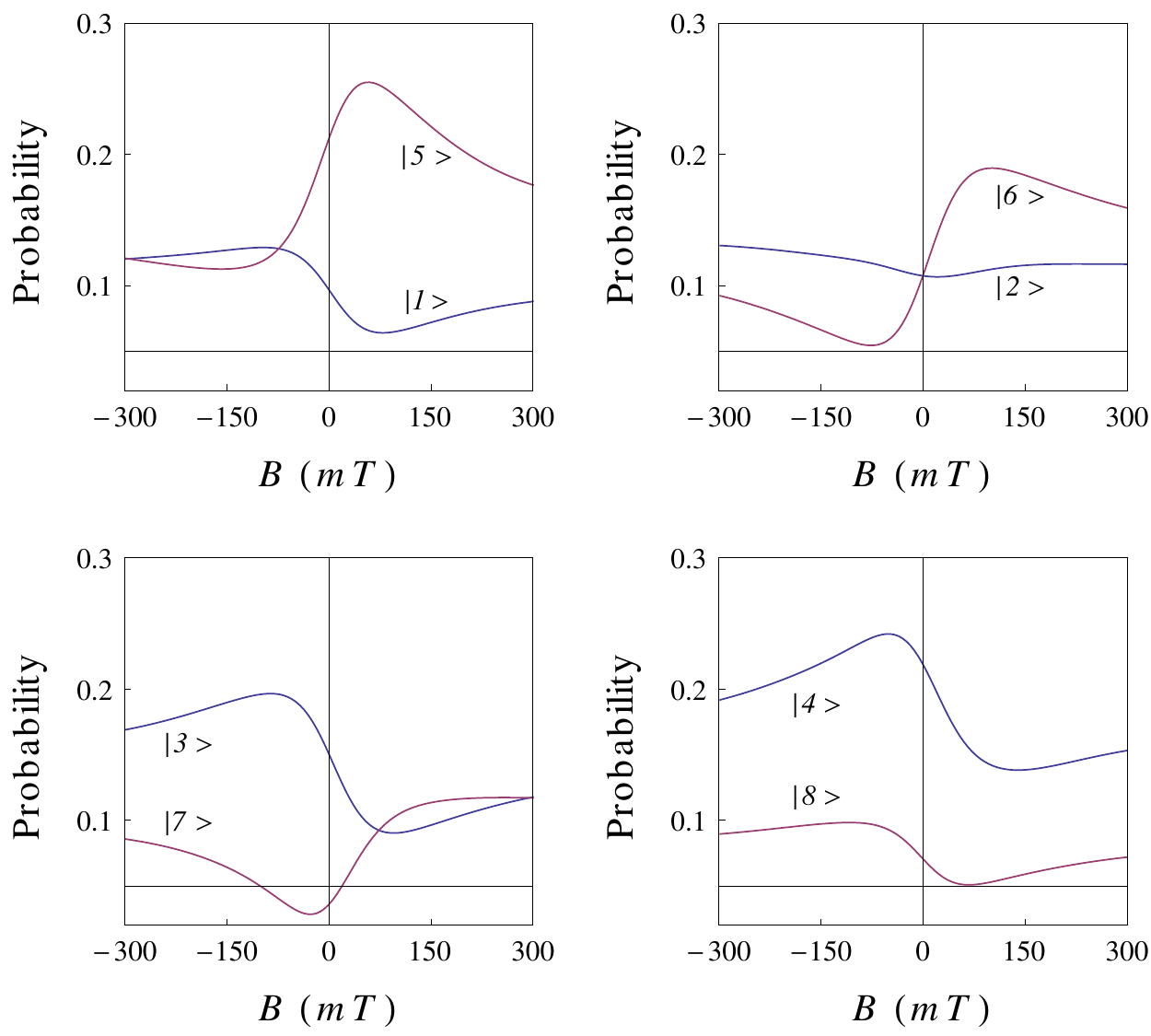}
\caption{(color on line) The calculated probability of the eight HFI-coupled states 
as a function of the magnetic field at $20$ mW pump power and right circularly polarized light. For $B$=0 the states corresponds to the ones listed in Eqs. (9) to (16).
}
\label{probs}
\end{figure}
\begin{figure}
\includegraphics[width=8.00 cm]{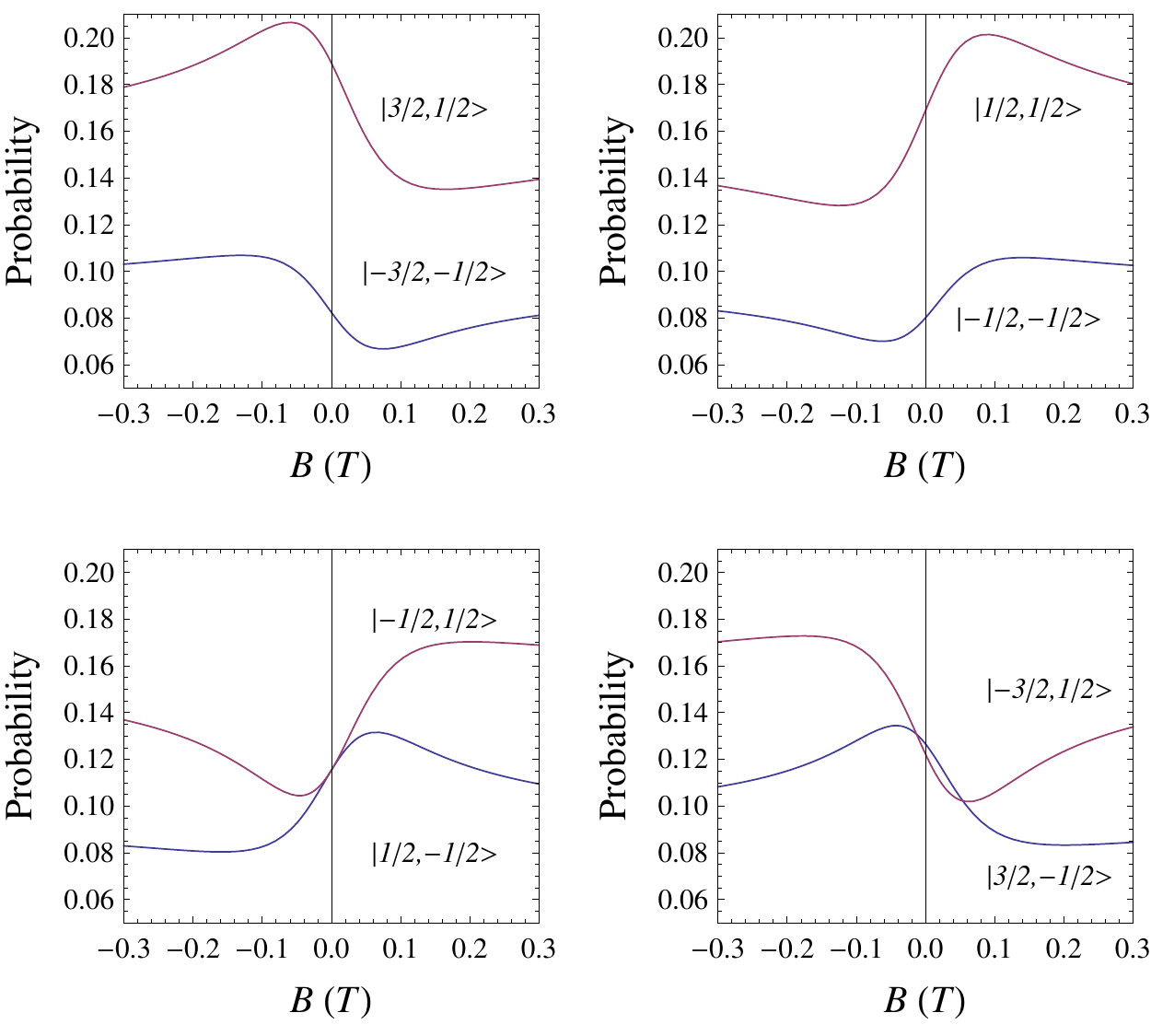}
\caption{(color on line) The calculated nuclear and localized electron
spin states probabilities as a function of the magnetic
field at $20$ mW pump power and right circularly polarized light.
}
\label{probs_pure}
\end{figure}
\begin{figure}
\includegraphics[width=0.5\textwidth]{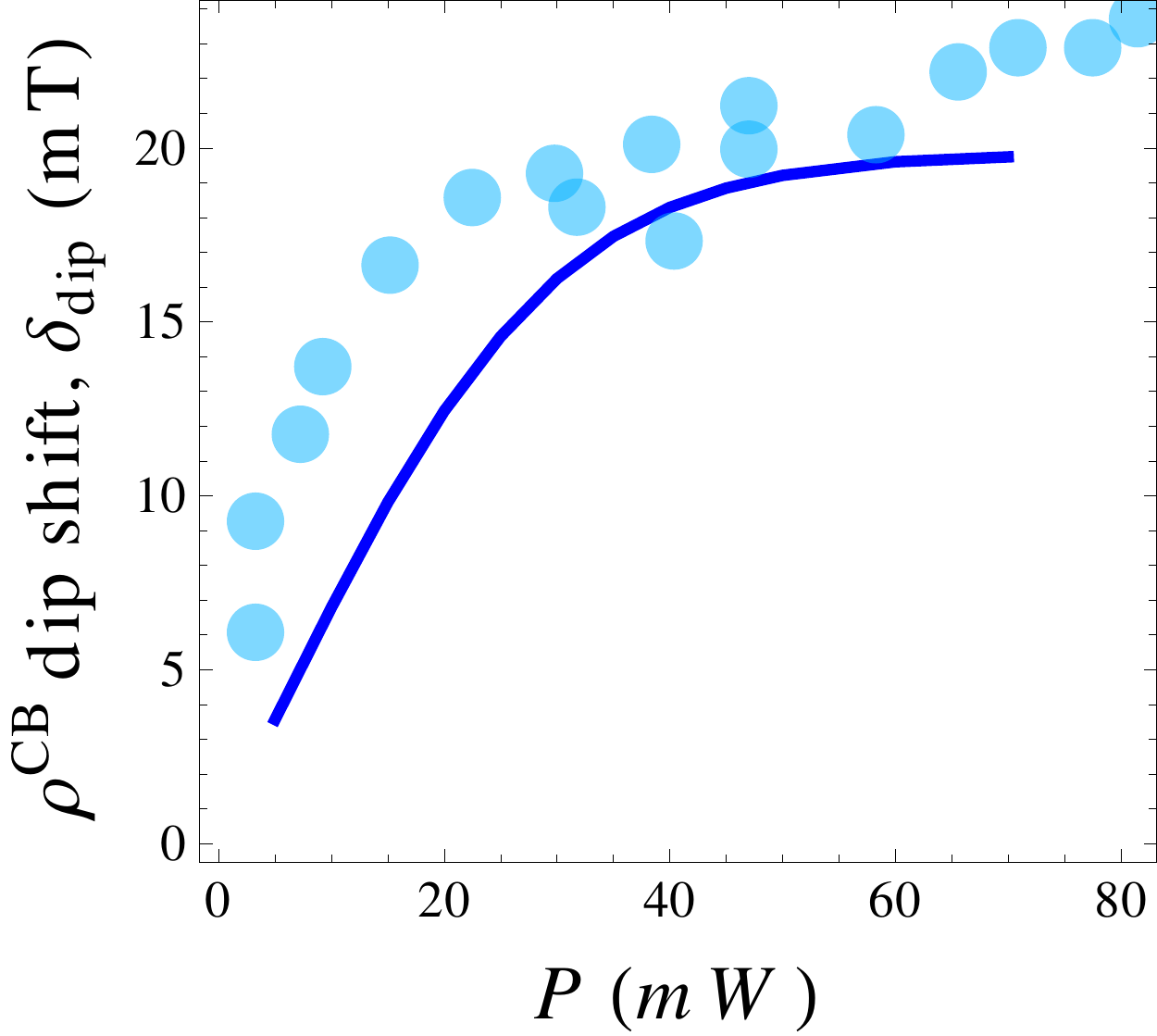}
\caption{Shift of the minimum of the conduction band polarization from the $B$=0 position as a function of the pump power $P$. The solid
line reproduces the theoretical calculation according to our model whereas  the dots indicate the
experimental points obtained by Kalevich, \textit{et al.}~\cite{kalevich:567} (see text).}
\label{beff}
\end{figure}
The 144 differential equations  arising from the master
equation (\ref{mastereq}) were
solved by fourth-order Runge-Kutta method.
Initially (before optical excitation) the localized electron-nuclear spin states are equally populated to
$N/8$ in order to guarantee zero
localized electron and nuclear spin polarization.
The rest of the variables and density matrix
elements were set to zero \textit{i.e.} we considered
unpopulated CB electron, VB hole and paired trap singlet states.\\
In Fig. 1 (a) the calculated SDR ratio under circularly polarized light is compared to the measured one. 
Values of the spin relaxation time of free and unpaired electrons on the centers
$\tau_{s}$=180 ps and $\tau_{sc}$=2200 ps respectively and
the effective hole life time $\tau_{h}$=13 ps
as well as the
typical ratio of the electron to the hole recombination coefficients
$\gamma_{e}/\gamma_{h}=6$ ( where $\gamma_{h}=1/\tau_{h}N$) are estimated from
previous time resolved PL experiments \cite{lagarde:208,kalevich:455}.
The ratio of bimolecular and hole recombination coefficients is set to
$\gamma_{r}/\gamma_{h}=0.008$. 
The calculated curve reproduces well the main features of the SDR power dependence: in low pumping regime the SDR$_r$ grows monotonically
until it reaches its maximum and finally, in the
strong pumping regime, it decreases monotonically. In the low pumping regime this
behavior has been attributed to the growing number of traps that dynamically
spin polarize in the same direction as the spin of the majority photo generated CB electrons
therefore augmenting their spin filtering effect. On the
contrary,  in the high pumping regime there is a large number
of photo-generated CB electrons compared to the total number of
centers. The CB electrons that are spin polarized in a
direction antiparallel to the traps dynamically
depolarize the latter thus reducing the spin filtering effect.
In addition, non-spin dependent recombination channels, such as bimolecular recombination itself, might
be present.
The model also describes very satisfactorily the SDR magnetic field dependence. Whereas for zero magnetic field the maximum
SDR$_r$ reaches approximately 225\%,  it increases up to SDR$_r$=260\% for $B$=185 mT.
Here the magnetic field seems to stabilize the localized electron spin polarization.
For magnetic fields above 200 mT the
SDR$_r$ saturates and remains constant.
Whereas the photoluminescence intensity for linearly polarized
light remains constant for all values of the magnetic field,
it is enhanced for larger magnetic fields
under circularly polarized light (not shown in the figure).\\
To gain further insight into the mechanism behind the amplification of the
spin filtering effect under magnetic field, we theoretically calculate the spin polarization degree
for CB electrons, localized electron and coupled nucleus for 
different pump power values using the parameters' values reported
in Ref. [\!\!~\citenum{kalevich:035205}]: $\tau_s=140 $ps, $\tau_{sc}=2200$ps, $\tau_h=30$ps,
$\gamma_e/\gamma_h=30$ and $\gamma_r=0$.
In Fig. \ref{figure4} (a) we observe the CB electrons spin
polarization degree $\rho^{CB}=2 S_z/n$ as a function
of the magnetic field in Faraday configuration for different
laser irradiances. Two main features can be evidenced: first, the amplification of the spin filtering effect as $B$ increases. Second, the shift of the CB spin polarization dip
from $B$=0 T. Concerning the first feature, we observe the same trend as Kalevich
\textit{et al.}~\cite{kalevich:035205}: the spin polarization degree increases
from its minimum $\rho^{\rm min}_s$ up to a saturation value $\rho^{\rm max}_s$
as the Faraday magnetic field absolute value grows. The
difference between these to extrema $\Delta P_s=P^{\rm max}_s-P^{\rm min}_s$
reaches a maximum at a pump power $P=20$ mW. As expected, the spin polarization degree of the localized electrons  $\rho^{loc}=2S_{cz}/N_1$ in Fig. \ref{figure4}
(b) follows a similar trend as
they are dynamically spin polarized by the CB electrons.
However, in Fig. \ref{figure4} (c) we observe a maximum for
the nuclear spin polarization degree $\rho^N=2J_z/3N_1$ aligned
on the same magnetic field value
as the minimum of the spin polarization degree of CB and localized electrons.
As the magnetic field in Faraday configuration is increased, the polarization of the
nuclear spin decreases until it vanishes at approximately $250$ mT.
The inflection point of the CB electrons, localized electrons and nuclei
spin degree of polarization
occurs close to $B=A\hbar/g_c\mu_B\sim 80$ mT where
the Zeeman energy and the HFI are comparable in magnitude.
From the previous analysis we can describe the effect of a Faraday magnetic field on the
spin filtering mechanism as follows:  Incident circularly polarized light spin pumps
CB electrons. Under strong magnetic field (such that the electron Zeeman interaction dominates over the hyperfine one), 
the localized electron is decoupled form the Ga$^{2+}_i$ nucleus and the eigenstates of the Hamiltonian are pure electron spin states.
During the recombination process of CB electrons into the traps, the resident electrons are dynamically spin polarized in the same
direction as the incoming CB electrons to the maximum degree possible under the actual excitation conditions. The interaction with the Ga$^{2+}_i$ nuclei is negligible (due to the strong Zeeman effect), and the nuclei retain their zero average angular momentum.  
However, for zero or weak magnetic fields (such that now the hyperfine interaction dominates over the electron Zeeman one),  
the hyperfine interaction mixes the localized electron and Ga$^{2+}_i$ spin states. On one hand the efficiency of the spin filtering mechanism is weakened
compared to a pure electron spin situation due to the partial lifting of the Pauli spin blockade. On the other hand, the same hyperfine interaction is responsible for the transfer of angular momentum to the nucleus leading to an increase of the nuclear spin polarization.\\
The second feature evidenced in Fig. 3 (a) is the 
asymmetric dependence of the average electron and nuclear polarization on the applied magnetic field direction for an excitation of a given helicity~\cite{kalevich:567}.
This feature is observable for nuclei with angular momentum larger than $1/2$, as it is the case here for Ga$^{2+}_i$ interstitial.
This shift from $B$=0 of the minimum of the conduction band electron spin polarization reflects an equal shift of the electrons spin polarization localized in the paramagnetic centers. 
These shifts arise due to the dynamical equilibrium under optical pumping with circular polarization due to the eigenstates' populations imbalance of the electron-nuclear system (see Figs.~\ref{probs},\ref{probs_pure}) compared to a uniform population condition. The increase of this shift with an increase of the excitation power reflects the modification of this dynamical equilibrium, the shift eventually saturating at a given value.
We emphasize here that this power dependent asymmetry is only obtained under dynamical equilibrium conditions. Although this behavior closely resemble an Overhauser effect, we clearly see that the concept of an effective nuclear magnetic field is not applicable in this context of a localized electron on a Ga nucleus in a strong coupling regime.\\
The calculated value of the conduction band polarization dip shift $\delta_{dip}$ from $B$=0  
 is plotted as a function
of the pump power in Fig. \ref{beff}. The dots indicate
the experimental results obtained by Kalevich \textit{et al.}~\cite{kalevich:567}
and the solid line are the theoretical results calculated from
the displacements simulated in Fig. (\ref{figure4}) (a) with the parameters corresponding to our own experimental results. We notice
that $\delta_{dip}$ increases until it saturates in the high power regime as
the nuclei acquire their maximum spin polarization and saturate.
This estimation is consistent with the experimental results.
The nuclear spin polarization maximum is forced to displace exactly to
the same value of $\delta_{dip}$ as the  minimum of the CB and bounded
electrons degree of spin polarization in order to ensure
spin transfer conservation, an essential characteristic of hyperfine interaction
[see Eqs. (\ref{momcons:eq1}) and (\ref{momcons:eq2})].
Thanks to this mechanism, it is possible to access the nuclear and localized electron spin polarizations from a measurement of the PL (or photoconductivity) polarization degree.  
As previously stated, under the influence of a magnetic field in Faraday configuration,
strong enough to make the Zeeman and HFI energies
comparable, the coupling between bounded electrons and nuclei becomes less efficient and
the spin mixing is lifted inhibiting  the transfer between bounded
electrons and nuclei.
The spin mixing occurring close to $B$=0 T is very large
between the nuclear-bounded electron states
$\left\vert 1/2,-1/2\right\rangle$ and
$\left\vert -1/2,1/2\right\rangle$ and
between states
$\left\vert 3/2,-1/2\right\rangle$ and
$\left\vert -3/2,1/2\right\rangle$ as can be seen
in Fig. \ref{probs_pure}. As the
magnetic field in Faraday configuration is increased this mixing
is lifted obtaining higher probabilities for those states
having positive bounded electron angular momentum for a $\sigma^+$ excitation.
The remaining spin states do not vary
considerably giving thus an overall increase in the bounded electron
spin polarization and an overall decrease in the nuclear spin polarization.
\begin{figure}
\includegraphics[width=8.00 cm]{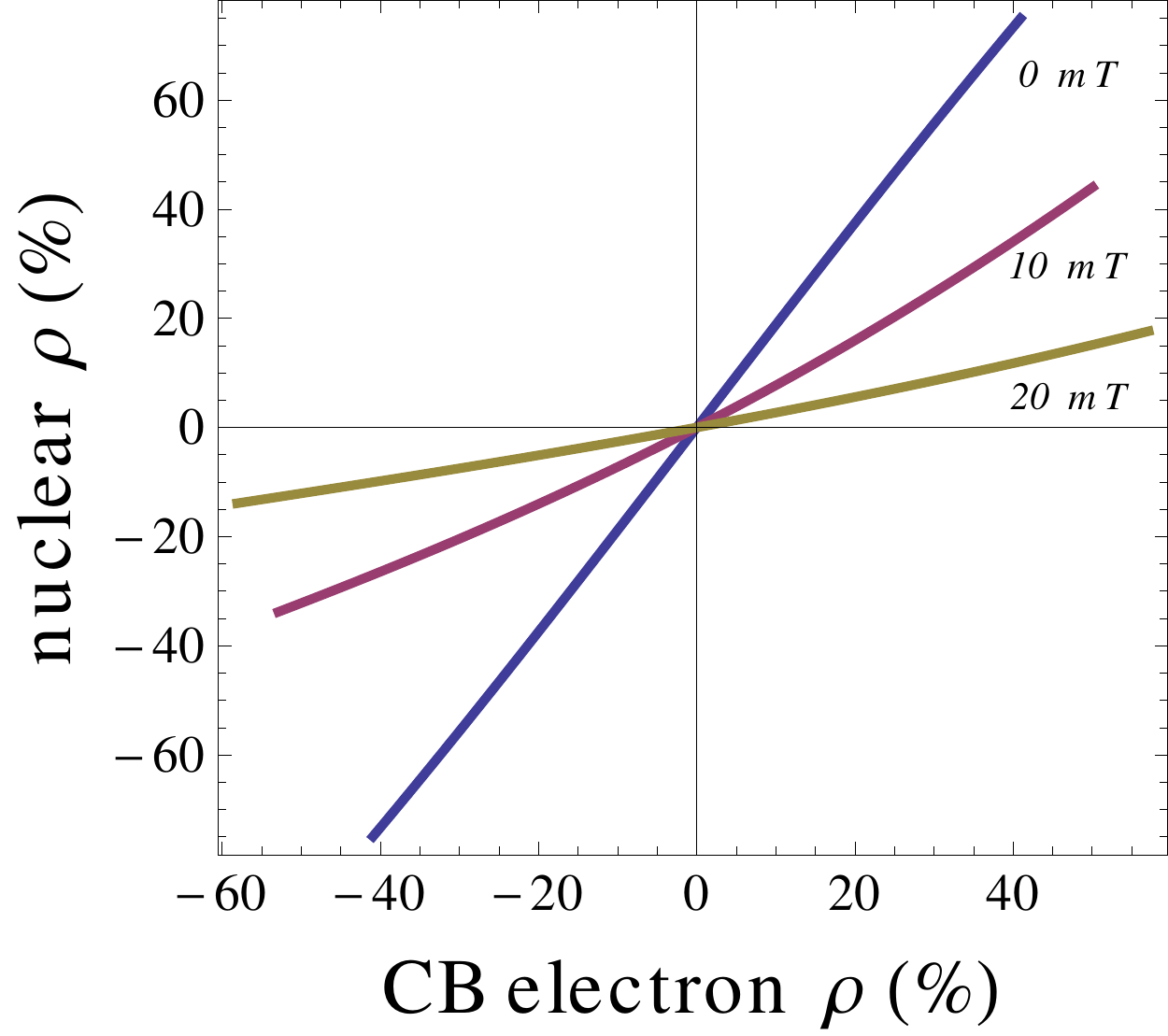}
\caption{(color on line) The calculated nuclear spin polarization degree as a function of CB electron spin polarization degree for $40$ mW 
pump power and various
values of the external magnetic field.
}
\label{figure7}
\end{figure}
Fig. \ref{figure7} presents 
the nuclear spin polarization
degree $\rho^N=2J_z/N_1$ as a function of the CB electron spin polarization
degree $\rho^{CB}=2S_z/N_1$ for various values of the
magnetic field shows a linear behavior. As expected, larger values
of the magnetic field impede nuclear spins to polarize due to
spin state mixing with the bounded electrons.

\section{Conclusions}\label{conclusions}
In summary we have demonstrated the possibility to access the
nuclear spin states of an ensemble of Ga$^{2+}_i$ centers through
a measurement of the optical polarization
of CB electrons in in GaAsN.
Optically spin pumped CB electrons dynamically polarize the
localized electrons in Ga$^{2+}_i$ centers by
spin dependently recombining with them. The
spin polarized localized electrons loose
angular momentum
to their corresponding nuclei by the hyperfine interaction. 
A control of the degree of the spin polarization of the Ga$^{2+}_i$ nuclei via the dynamical
polarization of CB electrons is thus possible. Our calculations show that the
nuclear spin polarization might be tuned with different excitation
parameters such as
pump power, degree of circular polarization of incident
light, and the intensity of a magnetic field in Faraday configuration.

The model  developed here describes all the essential features of the experimental
results as the dependence of SDR$_r$ and the spin polarization degree
as a function of the magnetic field and laser power.
It is capable of reproducing the shift of the CB electron spin
polarization degree curves as a function of the magnetic field
in Faraday configuration \cite{kalevich:035205,kalevich:567}.
 This feature is shown to be
caused by a dynamical equilibrium of the populations of the electron-nuclear states, strongly coupled via the hyperfine interaction, in the traps driven the spin dependent recombination.

\acknowledgments
Part of this work has been done in the framework of the EU Cost action N$^\circ$ MP0805.
Alejandro Kunold acknowledges financial support from
UAM-A CB department and thanks INSA-Toulouse for a two-months professorship position.

\appendix
\section*{Matrix representation of the operators}\label{operators}
Here we present the matrix representation of the
number and spin operators needed to build the
master equation. They are all $12\times 12$ matrices written in the basis:
\begin{equation}
\begin{split}
\mathcal{B} =\{
\left\vert 1 \right\rangle=\left\vert h \right\rangle, \left\vert 2 \right\rangle=\left\vert \mathrm{singlet} \right\rangle,
\left\vert 3 \right\rangle=\left\vert c\downarrow \right\rangle, 
\left\vert 4 \right\rangle=\left\vert c \uparrow \right\rangle,  \\
 \left\vert 5 \right\rangle=\left\vert -\frac{3}{2} \downarrow \right\rangle,
\left\vert 6 \right\rangle=\left\vert -\frac{1}{2} \downarrow \right\rangle,
\left\vert 7 \right\rangle=\left\vert \frac{1}{2} \downarrow \right\rangle, 
\left\vert 8 \right\rangle=\left\vert \frac{3}{2} \downarrow \right\rangle, \\
 \left\vert 9 \right\rangle=\left\vert -\frac{3}{2} \uparrow \right\rangle,
\left\vert 10 \right\rangle=\left\vert -\frac{1}{2} \uparrow \right\rangle,
\left\vert 11 \right\rangle=\left\vert \frac{1}{2} \uparrow \right\rangle,
\left\vert 12 \right\rangle=\left\vert \frac{3}{2} \uparrow \right\rangle\}
\end{split}
\end{equation}
where states from 1 to 4 represent respectively  the valence band hole, the  paired localized electron singlet state and the conduction band electron with their spin represented by the arrows. States from 5 to 12 each represents  a state of a given projection of the nuclear and localized electron spins.

The VB hole and CB electron number operators are
given by
\begin{eqnarray}
\left(\hat p\right)_{ij}
  &=&\delta_{i,1}\delta_{j,1},\\
\left(\hat n\right)_{ij}
  &=& \delta_{i,3}\delta_{j,3}+\delta_{i,4}\delta_{j,4},
\end{eqnarray}
The unpaired and paired trap number operators can
be expressed as
\begin{eqnarray}
\left(\hat N_1\right)_{ij}
  &=&\sum_{k=1,8}\delta_{i,k+4}\delta_{j,k+4},\\
\left(\hat N_2\right)_{ij}
  &=&\delta_{i,2}\delta_{j,2}.
\end{eqnarray}
The CB electron spin operators are given by
\begin{eqnarray}
\left(\hat{S}_x\right)_{ij}
  &=& \frac{1}{2}\left(\delta_{i,3}\delta_{j,4}
  +\delta_{i,4}\delta_{j,3}\right),\\
  \left(\hat{S}_y\right)_{ij}
  &=&\frac{i}{2}\left(\delta_{i,3}\delta_{j,4}
  -\delta_{i,4}\delta_{j,3}\right),\\
\left(\hat{S}_z\right)_{ij}
  &=& \frac{1}{2}\left(-\delta_{i,3}\delta_{j,3}
  +\delta_{i,4}\delta_{j,4}\right).
\end{eqnarray}
The nuclear spin operators are given by
\begin{eqnarray}
\left(I_x\right)_{ij}
&=&
\frac{\sqrt{3}}{2}\left(
\delta_{i,5}\delta_{j,6}+\delta_{i,6}\delta_{j,5}
+\delta_{i,7}\delta_{j,8}+\delta_{i,8}\delta_{j,7}\right.\nonumber\\
&&\left.+\delta_{i,9}\delta_{j,10}+\delta_{i,10}\delta_{j,9}
+\delta_{i,11}\delta_{j,12}+\delta_{i,12}\delta_{j,11}
\right)\nonumber\\
&&+\left(\delta_{i,6}\delta_{j,7}+\delta_{i,7}\delta_{j,6}
+\delta_{i,10}\delta_{j,11}+\delta_{i,11}\delta_{j,10}\right),\nonumber\\
\\
\left(I_y\right)_{ij}
&=&
i\frac{\sqrt{3}}{2}\left(
\delta_{i,5}\delta_{j,6}-\delta_{i,6}\delta_{j,5}
+\delta_{i,7}\delta_{j,8}-\delta_{i,8}\delta_{j,7}\right.\nonumber\\
&&\left.+\delta_{i,9}\delta_{j,10}-\delta_{i,10}\delta_{j,9}
+\delta_{i,11}\delta_{j,12}-\delta_{i,12}\delta_{j,11}
\right)\nonumber\\
&&+\left(\delta_{i,6}\delta_{j,7}-\delta_{i,7}\delta_{j,6}
+\delta_{i,10}\delta_{j,11}-\delta_{i,11}\delta_{j,10}\right),\nonumber\\
\\
\left(I_z\right)_{ij}
&=&\frac{3}{2}\left(
-\delta_{i,5}\delta_{j,5}+\delta_{i,8}\delta_{j,8}
-\delta_{i,9}\delta_{j,9}+\delta_{i,12}\delta_{j,12}
\right)\nonumber\\
&&+\frac{1}{2}\left(
-\delta_{i,6}\delta_{j,6}+\delta_{i,7}\delta_{j,7}
-\delta_{i,10}\delta_{j,10}+\delta_{i,11}\delta_{j,11}
\right)\nonumber\\
\end{eqnarray}
The auxiliary operators are useful in expressing
the dissipator and some of the other operators.
They can be expressed as
\begin{equation}
\hat{\vec{\sigma}}_m=\left(\hat{\sigma_{x,m}}, \hat{\sigma_{y,m}},\hat{\sigma_{z,m}}\right)
\end{equation}
with the definitions
\begin{eqnarray}
\left(\sopel_{x,-\frac{3}{2}}\right)_{ij}
  &=&\frac{1}{2}\left(-\delta_{i,5}\delta_{j,9}
     +\delta_{i,9}\delta_{j,5}\right), \\
\left(\sopel_{x,-\frac{1}{2}}\right)_{ij}
  &=& \frac{1}{2}\left(-\delta_{i,6}\delta_{j,10}
     +\delta_{i,10}\delta_{j,6}\right),\\
\left(\sopel_{x,\frac{1}{2}}\right)_{ij}
  &=& \frac{1}{2}\left(-\delta_{i,7}\delta_{j,11}
    +\delta_{i,11}\delta_{j,7}\right), \\
\left(\sopel_{x,\frac{3}{2}}\right)_{ij}
  &=& \frac{1}{2}\left(-\delta_{i,8}\delta_{j,12}
    +\delta_{i,12}\delta_{j,8}\right).
\end{eqnarray}
\begin{eqnarray}
\left(\sopel_{y,-\frac{3}{2}}\right)_{ij}
  &=&\frac{i}{2}\left(\delta_{i,5}\delta_{j,9}
     +\delta_{i,9}\delta_{j,5}\right), \\
\left(\sopel_{y,-\frac{1}{2}}\right)_{ij}
  &=& \frac{i}{2}\left(\delta_{i,6}\delta_{j,10}
     +\delta_{i,10}\delta_{j,6}\right),\\
\left(\sopel_{y,\frac{1}{2}}\right)_{ij}
  &=& \frac{i}{2}\left(\delta_{i,7}\delta_{j,11}
    +\delta_{i,11}\delta_{j,7}\right), \\
\left(\sopel_{y,\frac{3}{2}}\right)_{ij}
  &=& \frac{i}{2}\left(\delta_{i,8}\delta_{j,12}
    +\delta_{i,12}\delta_{j,8}\right),
\end{eqnarray}
\begin{eqnarray}
\left(\sopel_{z,-\frac{3}{2}}\right)_{ij}
  &=&\frac{1}{2}\left(-\delta_{i,5}\delta_{j,5}
     +\delta_{i,9}\delta_{j,9}\right), \\
\left(\sopel_{z,-\frac{1}{2}}\right)_{ij}
  &=& \frac{1}{2}\left(-\delta_{i,6}\delta_{j,6}
     +\delta_{i,10}\delta_{j,10}\right),\\
\left(\sopel_{z,\frac{1}{2}}\right)_{ij}
  &=& \frac{1}{2}\left(-\delta_{i,7}\delta_{j,7}
    +\delta_{i,11}\delta_{j,11}\right), \\
\left(\sopel_{z,\frac{3}{2}}\right)_{ij}
  &=& \frac{1}{2}\left(-\delta_{i,8}\delta_{j,8}
    +\delta_{i,12}\delta_{j,12}\right),
\end{eqnarray}
The number operators of nuclear state $m=\left\{-3/2, -1/2,1/2,3/2\right\}$ are:
\begin{eqnarray}
\left(\iop_{-\frac{3}{2}}\right)_{ij}
  &=&\delta_{i,5}\delta_{j,5}
     +\delta_{i,9}\delta_{j,9}, \\
\left(\iop_{-\frac{1}{2}}\right)_{ij}
  &=& \delta_{i,6}\delta_{j,6}
     +\delta_{i,10}\delta_{j,10},\\
\left(\iop_{\frac{1}{2}}\right)_{ij}
  &=& \delta_{i,7}\delta_{j,7}
    +\delta_{i,11}\delta_{j,11}, \\
\left(\iop_{\frac{3}{2}}\right)_{ij}
  &=& \delta_{i,8}\delta_{j,8}
    +\delta_{i,12}\delta_{j,12},
\end{eqnarray}
The remaining operators can be expressed in terms
of the auxiliary operators. The localized electron
spin operators are then given by
\begin{equation}
\hat{\boldsymbol{S}}_c=\sum_{m=-3/2}^{3/2}\sop_{m},
\end{equation}
and the nuclear angular momentum operator:
\begin{equation}
\hat{I}_z=\sum_{m=-3/2}^{3/2}m\iop_{m}.
\end{equation}
The generation term is given by
\begin{equation}
\hat{G} =
G\hat p+\frac{G}{2}\hat n+(G_+-G_-)\hat S_z,
\end{equation}
where the first two terms account for the
photogeneration of VB holes and CB electrons and the
last for the generation of spin polarization in the sample.

\bibliography{kunold}

\end{document}